\documentclass[sn-mathphys,Numbered]{sn-jnl}% Math and Physical Sciences Reference Style

\usepackage{graphicx}%
\usepackage{multirow}%
\usepackage{amsmath,amssymb,amsfonts}%
\usepackage{amsthm}%
\usepackage{mathrsfs}%
\usepackage[title]{appendix}%
\usepackage{xcolor}%
\usepackage{textcomp}%
\usepackage{manyfoot}%
\usepackage{booktabs}%
\usepackage{algorithm}%
\usepackage{algorithmicx}%
\usepackage{algpseudocode}%
\usepackage{listings}%
\usepackage{adjustbox}

\usepackage[T1]{fontenc}
\usepackage[utf8]{inputenc}

\begin{document}

\title[Citizen science for social physics]{Citizen science for social physics: Digital tools and participation}

\author*[1,2]{\fnm{Josep} \sur{Perell\'o}}\email{josep.perello@ub.edu}

\author[1,2]{\fnm{Ferran} \sur{Larroya}}\email{ferran.larroya@ub.edu}

\author[1,2]{\fnm{Isabelle} \sur{Bonhoure}}\email{isabelle.bonhoure@ub.edu}

\author[1,2]{\fnm{Franziska} \sur{Peter}}\email{fpeter@ub.edu}


\affil*[1]{\orgdiv{OpenSystems Research Group, Departament de F\'isica de la Mat\`eria Condensada}, \orgname{Universitat de Barcelona}, \orgaddress{\street{Mart\'i i Franqu\`es, 1}, \city{Barcelona}, \postcode{08028}, \country{Spain}}}

\affil[2]{\orgdiv{Universitat de Barcelona Institute of Complex Systems UBICS}, \orgaddress{\street{Mart\'i i Franqu\`es, 1}, \city{Barcelona}, \postcode{08028}, \country{Spain}}}

\abstract{Social physics is an active and diverse field in which many scientists with formal training in physics study a broad class of complex social phenomena. Social physics investigates societal problems but most often does not count on the active and conscious participation of the citizens. We here want to support the idea that citizen science, and more particularly citizen social science, can contribute to the broad field of social physics. We do so by sharing some of our own experiences during the last decade. We first describe several human mobility experiments in urban contexts with the participation of concerned young students, old women or other different groups of neighbours. We second share how we have studied community mental health care provision in collaboration with a civil society organisation and with the intense involvement of persons with lived experience in mental health. In both cases, we narrow down the discussion to digital tools being used and the involved participatory dynamics. In this way, we share key learnings to enhance a synergistic relationship between social physics and citizen science and with the aim increase the societal impact of the research on complex social phenomena.}



\maketitle

\section{Introduction}\label{sec1}

Human-behaviour research topics can today be grounded on a data-driven basis which was inconceivable only 20 years ago. Digital transformation of our societies with the broad use of the internet has led to a variety of digital interactive platforms and fully-equipped mobile phones. These devices can store extensively and intensively human movements, decision making processes or emotional reactions. They have opened the door to analyse and model both individual and aggregated human behaviours in an unprecedented manner.

This sort of data-driven research has deep interdisciplinary and multidisciplinary roots. Many of the academics contributing to a better understanding of the related social phenomena are physicists \cite{Jusup2022}, generally with a strong background in statistical physics \cite{Castellano2009,Wang2016,Perc2017,Barthelemy2019,Sanchez2018} and complex systems science \cite{Ball2012}. Physicists have seen the need of cutting across academic boundaries. Their curiosity has led them to look outside their traditional domains \cite{Stewart1947,Stewart1950} expecting to unveil laws similar to those in physics \cite{Perc2019}. Jusup and coauthors \cite{Jusup2022} define social physics “as a collection of active research topics aiming to resolve societal problems to which scientists with formal training in physics have contributed and continue to contribute substantially”. The authors also qualify social physics as an extremely active and diverse field which broadly includes human behavior and interaction, but also human cooperation or human mobility to just name a few of topics \cite{Jusup2022,Barbosa2018,Castellano2009,Wang2016,Sanchez2018,Perc2017,Barthelemy2019,Ball2012,Stewart1947,Stewart1950,Perc2019}.

Citizen science is also being a growing research practice during the last decade \cite{Irwin2018}. Citizen science broadly refers to the active engagement of the general public in scientific research tasks \cite{Vohland2021}. In citizen science, scientists and citizens collaborate to produce new knowledge for science and society. Crowdsourced data can be collected with Apps on mobile phones or web-based platforms. Amateurs have volunteered to take pictures of for instance bird species, invasive marine species or tiger mosquitoes. 

The outcomes of this joint effort is already showing relevant scientific impact \cite{Cooper2014}. In the particular field of physics, participants have classified millions of galaxies thanks to a well-documented website and a carefully guided interaction. The articulated effort has been key in numerous scientific publications \cite{Marshall2015}. In another level but still within the context of physics, local communities have created a network of low cost earthquake sensors \cite{Calais2022} or became key actors in the quick reaction after the Fukushima disaster collecting extensive radiation measurements \cite{Brown2016}. A scientific research can be thus organised by building collaborative networks of participants. These networks can also serve to effectively report volcano observations in Europe or eventually in all parts of the world \cite{Sandri2023}. All these cases are just few examples of the wide set of initiatives flourishing worldwide following very different formats and strategies.

However, it is not that frequent to find citizen science projects within the specific field of social physics. Human-related activities and human behavior topics make less clear the roles and the position of the participants in front of the social topic under investigation which very often is related to social concerns that are at stake \cite{Albert2021,Bonhoure2023}. There is a sophistication and a subtlety in the research that contrasts with more established citizen science practices on for instance biodiversity monitoring or astronomy observation.

We want here to further support the idea that citizen science practices can effectively bring everyone into the important task of better understanding those complex social phenomena \cite{Jusup2022}. To encourage researchers from social physics to further embrace citizen science, more reflection is needed. There are not enough spaces for sharing experiences from which extract key learnings. These spaces can help to further reflect in the configuration of the related digital tools which needs to be adaptive to participants' perspective, skills, concerns and motivation. The building or adaptation of digital tools for crowdsourced data collection is not the only one aspect to be considered in citizen science practices. To favour a citizen science for social physics, it is also necessary to further reflect on why and how citizens can be involved  and carefully structure participation in research activities. At least part of the reflection can be nurtured by the so-called citizen social science or social citizen science \cite{Albert2021,Bonhoure2023} which enhances social dimensions in citizen science projects and increments participatory research on social shared concerns with communities, sometimes in a vulnerable situation.

Our OpenSystems research group \cite{OpenSystems} has been developing during the last decade in urban contexts and mostly in the Barcelona (Spain) metropolitan area \cite{Bonhoure2023,OpenSystems, Gutierrez2014,Sagarra2016,Gutierrez2016,Gutierrez2016b,Poncela2016,Senabre2018,Vicens2018,Vicens2018b,Cigarini2018,Cigarini2020,Cigarini2021,Larroya2023,Perello2021,Perello2022}.  We will restrict the paper to the presentation of two distinct themes with the aim to describe in very practical terms the development and the implementation of digital tools to collect crowd-sourced data, and the ways and reasons to incorporate citizens' participation into the research activities described. The first theme is being traditionally extensively explored in the context of social physics: human mobility \cite{Jusup2022,Barbosa2018}. The second theme focuses on community mental health care provision and it shows its link to social interactions \cite{Castellano2009}, human cooperation \cite{Perc2017,Sanchez2018} and complex systems science in general \cite{Ball2012}. Discussion and conclusion section summarizes and reflects on main aspects presented.

\section{Human mobility}\label{sec2}

Human mobility has been gaining attention within the physics community \cite{Barbosa2018}. Contributions have analyzed location data to learn about existing patterns \cite{Gonzalez2008}. Some of them have also provided new models or, alternatively, improved or assessed the existing ones \cite{Gallotti2016,Simini2012,Chen2018}. Some of the papers have also brought out relevant insights about urban contextual issues like walkability or COVID response policies \cite{Rhoads2021,Hunter2021}. 

Census data has been for decades the (public) data source, the same for all the academic community. It has been able to assess the quality of well-known models in the literature such as the radiation model \cite{Simini2012}. However, census data might be unsatisfactory for the actual research agenda. Higher level of data granularity is very much often required, with higher time and space resolution. The wide use of mobile phones and their related technologies give the capacity to collect the location data \cite{Blondel2015}. Social media accesses/posts collected by the researchers themselves \cite{Llorente2016} or mobile phone call data records (CDR) provided to specific research groups are an alternative to gain knowledge on human mobility patterns at a mesoscopic level \cite{Schlapfer2021}. If it is of interest to look at an even higher resolution scale, then other tracking systems might be necessary. This can be for instance the case of specific mobile applications (Apps) tracking systems \cite{Moro2021}.

Highest-resolution mobility data is generally owned by private companies. However, in some cases, societal challenges and humanitarian purposes have motivated a closer collaboration with academia. A mobile company worked with academic researchers and the researchers analyzed the movements of 1.9 million mobile phone users around Port au Prince city before and after the Haiti earthquake in 2010 \cite{Lu2012}. Some companies opted to launch open calls for receiving specific academic demands (related to humanitarian purposes) and the companies selected an identified group of scientists to have access to very valuable location data \cite{Taylor2016}. Hackathons and similar gatherings had also been a strategy by telecommunications companies in Africa to mobilize the academic community towards sustainable or societal challenges by releasing partial data from some countries \cite{Blondel2015} but without questioning ethical and privacy related issues. 

All in all, it is still difficult to find open mobility data. Data reuse is thus representing an obstacle for scientific reproducibility and the further advancement of the related science \cite{Lazer2020}. Research agenda is thus mostly driven on which data a researcher have access to. In scientific papers, raw data is most often said to be not available to other researchers and only an anonymized and coarse grained data set might be offered upon request (see for instance the cases of Refs. \cite{Hunter2021,Gallotti2016,Schlapfer2021,Moro2021}). Some scientific papers alternatively recommend contacting the company owning the data if the reader wants to access and perform new research over the same data.

\subsection{Digital tools: crowdsourced data collection and human mobility}\label{digitaltools}

Concerns within human mobility research community are not limited to data accessibility and scientific reproducibility. There is also an intricate ethical debate that involves informed consent by the owners of the mobile phones. The debate indeed puts at the center of the discussion whether it is feasible to use highly sensitive data as a public good \cite{Taylor2016}. 

There are initiatives that have built the whole research around a digital tool avoiding the researchers’ intermediation with telecommunication companies \cite{Eagle2006}. This approach is deeply linked with citizen science practices and can overcome privacy and ethical problems, as participants are clearly informed and specifically consent to their data being used.

\subsubsection{Building and updating an App}

\begin{figure}
\centering
\includegraphics[width=0.455\textwidth]{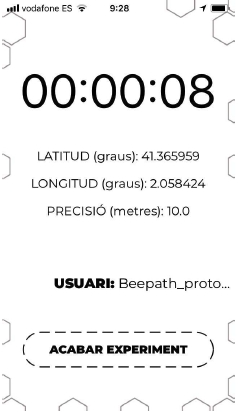} \qquad
\includegraphics[width=0.41\textwidth]{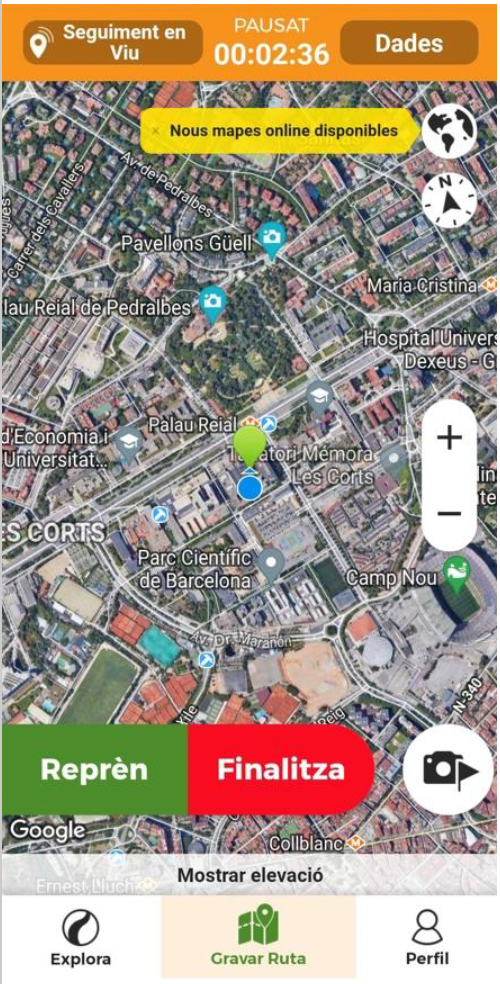}
\caption{{\bf Screenshots of the mobile Apps being used to collect GPS data.} Left image shows the information displayed by the BeePath App: time, latitude, longitude, GPS precision, username and a button to finish the experiment. Right image shows the information displayed by the Wikiloc App: time, a map with the trajectory, and two buttons to restart (or pause) and to finish the journey recording. Texts are in Catalan.}
\label{fig:App}
\end{figure}

Under this more open crowdsourcing basis, some of us and other collaborators started new citizen science research in human mobility by building from scratch an App called BeePath (see a screenshot in Figure \ref{fig:App}). The App worked in iOs and Android mobile phones and the first version was used in 2012 and left open in github \cite{gitbee-path}. We were able to collect high resolution GPS data every second and sent it to our university servers. The data was also shown in real time in an anonymized manner on a screen placed in a hotspot within the Barcelona Science Festival (Festa de la Ciència, in Catalan) located at Parc de la Ciutadella (4 ha, Barcelona). Visitors were recruited as participants of the mobility experiment. This action was part of the efforts to promote citizen science within the frame of the Barcelona Citizen Science Office \cite{Barcelona} (an initiative led by the Barcelona municipality). 

We updated and modified the BeePath App several times since then. Several other citizen science investigations developed were valuable to participants themselves but less relevant for academic audiences and scientific journals. In 2017, we finally decided to run a large-scale pedestrian mobility experiment in a broader level, in urban contexts around 10 schools in the Barcelona metropolitan area (see Subsection \ref{Sect:Purposed} below for further details). The data collected is carefully described and fully accessible in Ref. \cite{Larroya2023}. We considered pedestrian mobility in a neighborhood level, covering a typical distance range of hundreds of metres ($\approx 700$m) around the participating schools in the experiment. The typical journey duration was about 8 minutes \cite{Larroya2023}. 

This latter version allowed participants to take further ownership of data collected. We built some basic protocols that must be followed by the participants to deanonymize GPS datasets from early start. The data from this experiment was sent to our server and the participants could consult a brief report on their own data and download the data in a $.csv$ format file. The data collected had however some additional potential risks to identify participants as natural persons inferring their home addresses. The risk thus needed to be handled with extra work made by professional scientists and using geo-masking techniques ($k-$anonimity) \cite{Larroya2023}. Privacy protection was also the reason why we did not collect gender identity or any other socio-demographic trait. All these measures can become a limitation in the scientific outcomes expected from participatory mobility experiments.

\subsubsection{Reusing existing Apps}

The constant changes in mobile phones Apps technical requirements oblige to release periodic upgrades. Unfortunately, these improvements add new challenges to a citizen science research led by a university and/or any organisation with limited resources. Besides, funding schemes by research agencies are most often not aligned with the necessary stable funding to maintain a long-term vision in citizen science projects, even if the amount of money is small. These funding schemes even sometimes determine that the expenses to keep an App alive might not be eligible as these sort of activities are still not seen as part of the daily research activity. A possible way out is to build a partnership with specific companies. They generally have the capacity and the flexibility to dedicate a stable taskforce to develop the new code and the related upgrades. This was for instance the case of some of the BeePath initiatives mentioned above.

Alternatively, it is also possible to take advantage of existing Apps in the market which have quite different purposes (e.g. trekking or other outdoor sports activities). We used this strategy to run a new mobility experiment with very few resources available and to set up the related logistics very quickly. Before running the experiment, we checked the GPS precision of the App. We also tested the procedure of how data is collected and on how stopping times were treated by one specific App (Wikiloc \cite{wikiloc}) with reverse engineering. See a screenshot of the Wikiloc App in Figure \ref{fig:App}. This approach minimized the time needed to set up the digital infrastructure and the related taskforce but it required to redefine the privacy and consent protocols with care to comply data privacy regulations (GDPR, in the European case). For privacy reasons, we opted to offer the mobile devices to the participants but this in turn limited the amount of participants. The devices already had an account activated so that we can easily collect data after the experiment. These aspects might however limit the scope and goals of the experiment. 

\subsection{Participation: engaging people in pedestrian mobility experiments \label{Sect:engagemobility}}

Existing research papers in the literature without active and conscious participation show very large amount of records. For instance, in Ref. \cite{Moro2021}, 67.0 billion GPS records from 4.5 million unique smartphones (from Cuebiq company, anonymized location data from applications of opted-in users in an anonymized way) are used to explore how mobility patterns relate to economic segregation in large US cities. Similarly, Ref. \cite{Hunter2021} reports the GPS locations of 1.62 million users in 10 US metropolitan areas with the aim to analyse the impact of COVID-19 pandemic response measures on walking behaviour. 

Our own citizen science experiences \cite{Gutierrez2016,Larroya2023} reported in Table \ref{tab:mobility_experiments} lack the large amounts of GPS records owned by telecommunication or digital companies that other scientific publications use. However, citizen science mobility experiments brings the opportunity to narrow down the scope of the research and obtain only the data that is more capable to respond a predefined research question. We thus present below how to study purposed-based mobility and how to activate emergent uses of public space with a mobility experiment. Citizen science also offers the opportunity to work with specific communities so that specific socio-demographic groups can be identified beforehand. This sort of information can be precious to better understand mobility patterns.

Following subsections report our experience for purposed-based pedestrian mobility in urban contexts \cite{Gutierrez2016,Larroya2023} to take further inspiration about the kind of contributions that citizen science can bring into this field. Table \ref{tab:mobility_experiments} shows basic information from each of the citizen science experiments being developed in 4 different urban contexts. 

\subsubsection{Purposed-based pedestrian mobility}
\label{Sect:Purposed}

\begin{table}[t]
\centering
\caption{\label{tab:mobility_experiments}{\bf Pedestrian mobility citizen science experiments in different urban contexts.} We report the key term, the main research question and its most relevant scientific outcome of 4 mobility experiments. Columns refer to the context in which the experiments were organised, the digital tool being used (a mobile App collecting GPS locations), the profile of the participants, the number of participants involved, and the number of GPS records collected in each of the human mobility experiments.}
\begin{tabular}{lccccc}
\hline \hline
Context  & Digital tool & Participant's profile & N. part. & GPS data \\ \hline
Science Festival  & BeePath App & General public & 101 & 4,994  \\ 
\multicolumn{5}{l}{Key term: navigation}\\
\multicolumn{5}{l}{Question: How do the people move in a crowded environment with several hotspots?}\\
\multicolumn{5}{l}{Outcome: Model in terms of waiting time, jump length distribution, and spot's attractiveness}
\\
\\
\hline
Context  & Digital tool & Participant's profile & N. part. & GPS data \\ \hline
Community Center   & BeePath App & Old women & 10-15 & 45,136 \\
\multicolumn{5}{l}{Key term: walkability} \\
\multicolumn{5}{l}{Question: Which are the most pleasant routes to walk?}\\
\multicolumn{5}{l}{Outcome: Velocity and other mobility metrics correlated to granular context and walkability}
\\ 
\\
\hline
Context  & Digital tool & Participant's profile & N. part. & GPS data \\ \hline
Schools  & BeePath App & Teenagers & 458 & 41,053 \\
\multicolumn{5}{l}{Key term: walkability} \\
\multicolumn{5}{l}{Question: Which paths do the students take to go from home to school?}\\
\multicolumn{5}{l}{Outcome: Velocity and other mobility metrics correlated to granular context and walkability} 
\\ 
\\
\hline
Context  & Digital tool & Participant's profile & N. part. & GPS data \\ \hline
Neighborhood & Wikiloc App & Kids classgroups, teachers,  & 72 & 2,981 \\  &  & old persons, and families\\
\multicolumn{5}{l}{Key term: urban planning} \\
\multicolumn{5}{l}{Question: Which are the most liveable public spaces in the neighborhood?}\\
\multicolumn{5}{l}{Outcome: Mobility patterns and waiting time to identify in which spaces intervene}
\\
\hline \hline
\end{tabular}
\end{table}

\begin{figure}[t]
\centering
\includegraphics[width=1\textwidth]{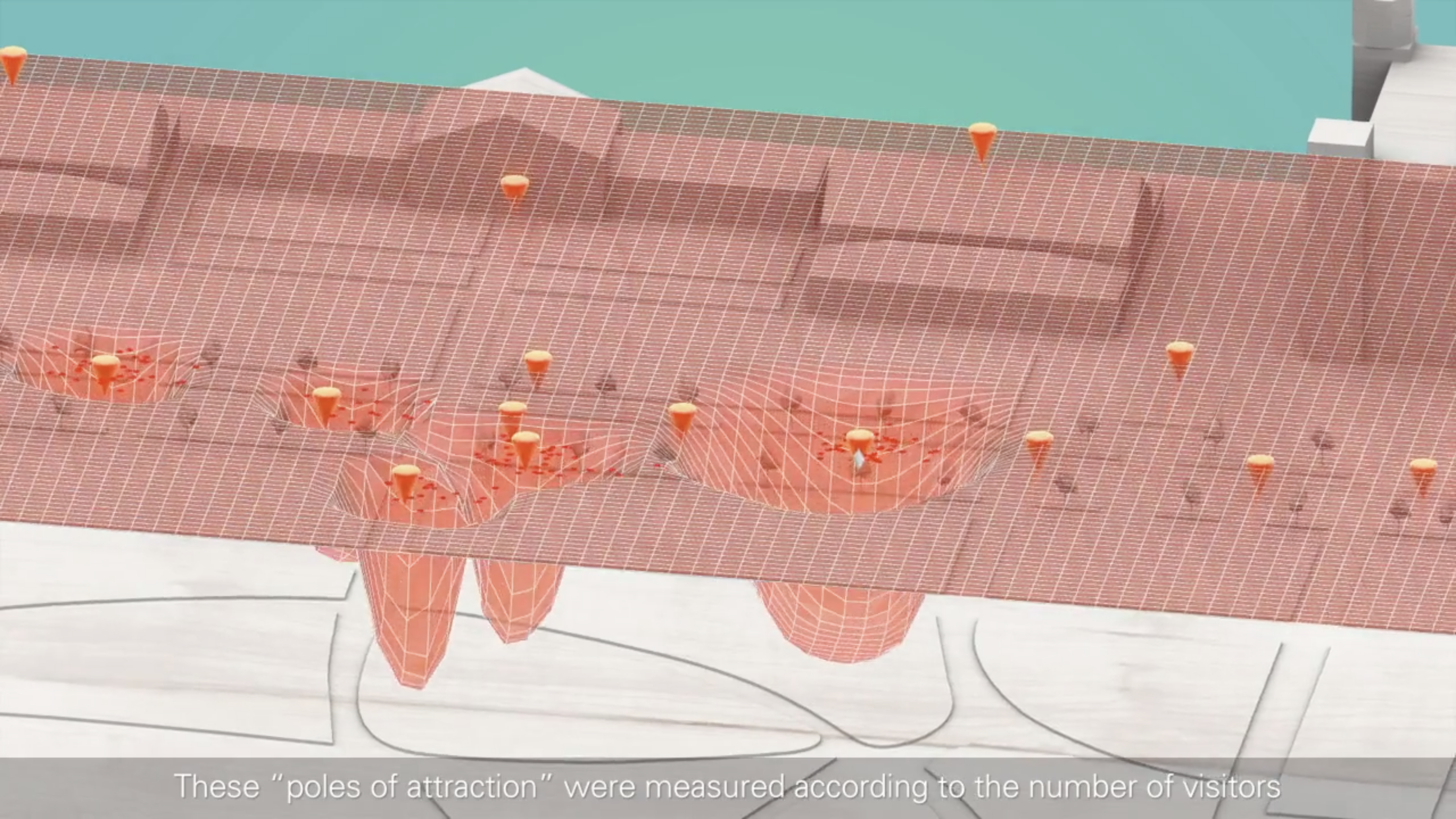}
\caption{{\bf Frame of the Barcelona Science Festival citizen science experiment promotional video.} We show the frame with a representation of the potential wells that qualitatively visualizes the attraction level of the spots inside Parc de la Ciutadella and where scientific outreach activities were taking place. The video can be found in Vimeo \cite{videofesta}.}
\label{fig:beepath_sciencefestival}
\end{figure}

In subsection \ref{digitaltools}, we already mentioned a very specific experimental setting within the Barcelona Science Festival and involving their visitors (101 participants). Scientific results are related to the exploration of the different spots in a public park and thus in a very well defined area. Active and reactive mobility patterns were identified \cite{Gutierrez2016}. Figure \ref{fig:beepath_sciencefestival} shows a frame of the video explaining the experiment \cite{videofesta}. We complied with data privacy regulations and signed the related informed consent on paper. We were also able to talk and discuss with interested citizens in participating to the experiment. We wanted to raise public awareness about the fact that our mobile phones are constantly collecting location data.

\begin{figure}
\centering
\includegraphics[width=1\textwidth]{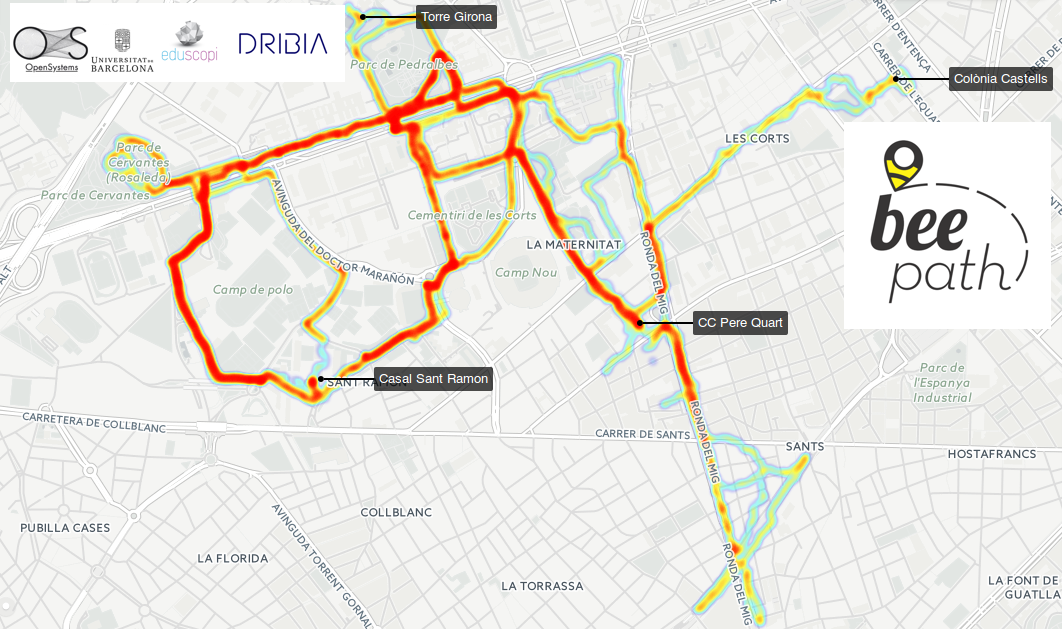}
\caption{{\bf Map visualization of the recurrent trajectories made by a group of old women from a community center.} We show the map being used for the discussions in the community center (Centre Cívic Pere Quart, CC Pere Quart) and being shared with the neighborhood to further promote walkability. Most suitable routes for the about 1h journeys were suggested. Departures and arrivalls were made from CC Pere Quart on Tuesdays and from another community center on Thursdays (Casal Sant Ramon). Two additional locations were added in the map as they were key targeted objectives. Beepath is a joint effort of OpenSystems-Universitat de Barcelona, Eduscopi (science education and communication company) and Dribia (data science company) and their logos appeared in these maps.}
\label{fig:beepath_centrecivic}
\end{figure}

Citizen science also brings the possibility to imagine repeated experiments with the same group. For healthy and well-being purposes, a group of about 10-15 old women used to come out together to walk for about an hour twice a week. The activity was part of a particular community center program (Centre C\'ivic Pere Quart, Les Corts, Barcelona). At least one member of the group activated the App in every journey. The journeys always started and ended at the two possible locations (depending on the day of the week). The women learn about new uses of their mobile phones. Also, the data collected served to identify most pleasant locations for a pedestrian (streets with green or tree shadows, wider side-walks...). The community center shared the routes taken by the group with the neighborhood under the form of maps (see one sample in Figure \ref{fig:beepath_centrecivic}). The joint effort appeared in the local tv and two old women were able to explain the main results of the research to the audience \cite{beteve}. Different scientific analysis were also made by one last-year graduate student in Physics from Universitat de Barcelona as part of his Bachelor's Degree Final Project.

Another experiment using the BeePath App involved teachers and students from 10 schools in the Barcelona metropolitan area. The effort was indeed considered an innovative STEAM activity in formal education. School participants were actively engaged and contributed to the project in different stages. They decided the scientific question about mobility: ``How do we arrive and leave the school?'', with the aim to study and learn how is the mobility in schools' vicinity and what obstacles students face in their daily home-to-school (or school-to-home) journey (e.g. narrow side-walks or absence of pedestrian crossings) \cite{Larroya2023}. 

In the different stages of the co-design process, participants developed and refined the experimental protocol for collecting data. They recorded their journey to/from school using an smartphone device with the BeePath App and in a predetermined time window. The students also became testers of the technology developed and interpreted their own school group's mobility data based on their knowledge of the school and the neighbourhood, and produced visualisations of the trajectories. The students finally presented the results in a public event in front of other school students and municipality representatives. The research done by the students resulted in the delivery of a set of evidence-based recommendations to the municipality representatives, in order to reach a wider urban perspective and thus improve accessibility to schools \cite{Diaz2023}. 

\begin{figure}
\centering
\includegraphics[width=1\textwidth]{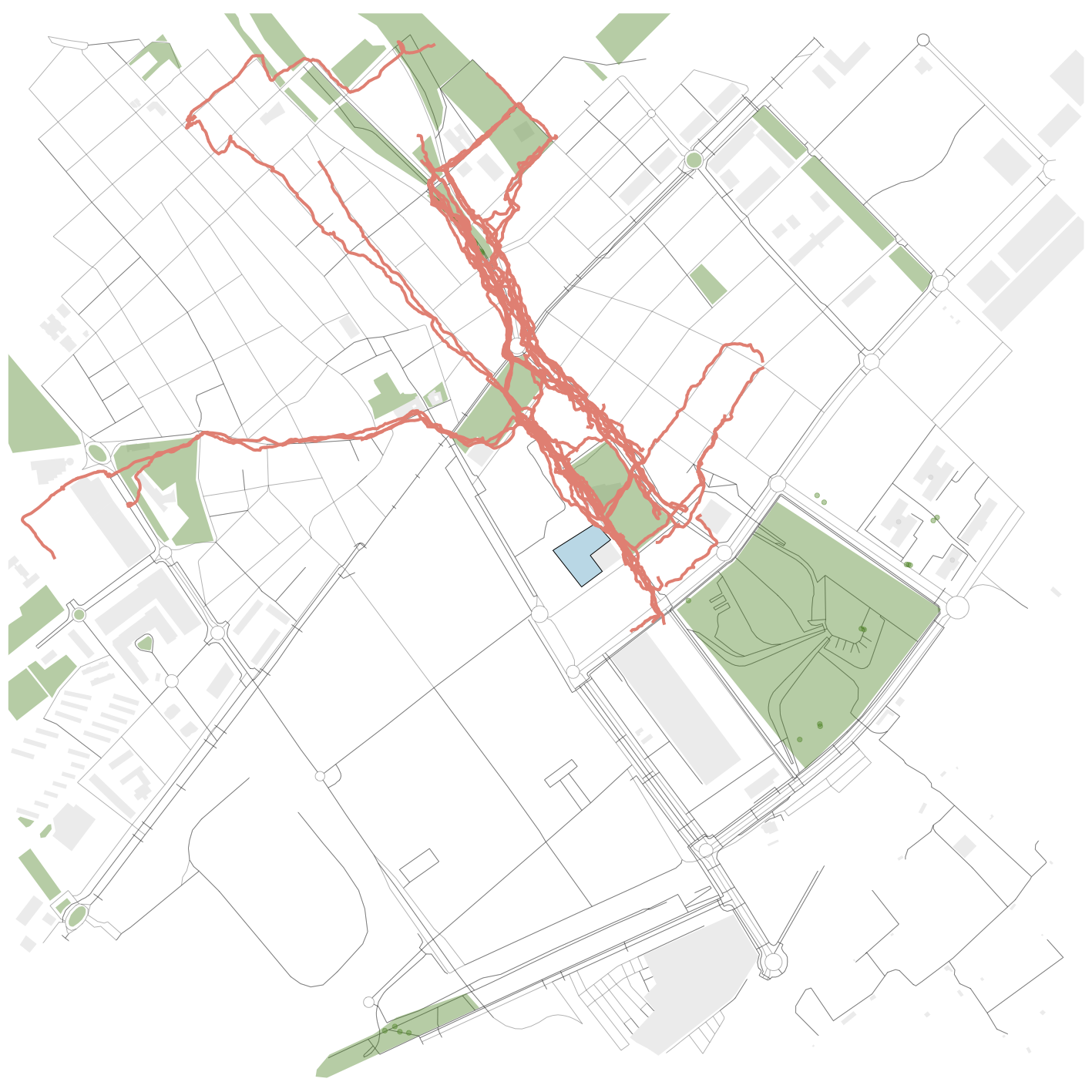}
\caption{{\bf Map visualization of the trajectories from BeePath in schools.} 28 processed home-to-school (or school-to-home) individual trajectories displayed (red) on a map around the Sant Gabriel de Viladecans school (blue), located at the city of Viladecans in the Barcelona metropolitan area. Data is geo-masked following the protocol described in Ref. \cite{Larroya2023}. Green areas correspond to public parks.}
\label{fig:beepath_school_map}
\end{figure}

A total of 427 Secondary school students (14-16 years old) and 31 teachers participated in the experiment. 262 out of the 427 students recorded the journey to (or from) school with their own smartphone. All of the details about the co-design and the co-creation phases of the experimental protocol, the data gathering, the data processing and filtering and the results obtained can be found in Ref. \cite{Larroya2023}. Figure \ref{fig:beepath_school_map} shares the visualisation of the trajectories to arrive or leave the Sant Gabriel de Viladecans school. In this case, 28 unique trajectories were gathered after filtering and processing the collected data. Different scientific analysis were also made by one last-year graduate student in Physics from Universitat de Barcelona as part of his Bachelor's Degree Final Project. Also, we will soon publish scientific results on walkability by means of different statistical analysis typically performed by physicists by means of magnitudes such as the instantaneous velocity and many related statistical features.

\subsubsection{Activating public spaces in a neighborhood through mobility}

To close this Section \ref{sec2}, we provide specific details of our most recent citizen science urban mobility experiment that does not focus on purposed-based mobility and sees pedestrian mobility and public participation as a way to imagine and discuss together possible transformations of the existing public space. The experiment was conducted in the ``Primer de Maig'' neighbourhood (3 ha, Granollers, Spain). The experiment was part of the research project ``Civic Placemaking: Design, Public Space and Social Cohesion". This project was led by architects and urbanists and some of us were invited to join the research. The project broadly seeked to promote social cohesion and inter-cultural integration through ephemeral architecture projects that have an impact on public space and rely on the active engagement of citizens \cite{Elisava2022}.

We there implemented citizen science participatory stategies and sophisticated methods for the data analysis that a physicist working on complex social phenomena would apply (e.g., waiting time distribution, transects lengths and reorientation angles or the identification of specific key locations with GPS data clustering). 72 people got involved in 19 groups. They became explorers of the neighbourhood, walking and seeking out specific places to perform a set of festive actions. The objective was to study the urban spaces of appropriation in an emergent manner, with a bottom-up approach.

The call for participation was launched through three local neighborhood associations. The engaged participants therefore had different socio-demographic profiles. Groups used tablet devices (one per group) with the Wikiloc App \cite{wikiloc} previously installed and registered with Google accounts by the research staff. In this way, participants did not have to provide any personal information that could compromise their anonymity. In addition, running the experiment in groups and starting data gathering from the public associations sites further guarantees the anonymity of the participants (for example, the location of their homes cannot be inferred). 

Data collection was divided into 2 consecutive days. 4 groups of teachers did the experiment in the morning and 7 groups of students (Primary school) in the afternoon. During the following day, the rest of the groups (8) participated to the experiment. The duration of the experiment and data collection was 1 hour and 30 minutes, in which participants explored the neighbourhood and completed 6 festive ``missions'' (e.g., choosing a place to open a drink and toast as a group, or finding a place to play a song on the speaker and dance).

The collected high-resolution pedestrian mobility data helped us to capture movement flows within the neighbourhood and to identify those sites chosen by participants to perform the actions (and their duration). Actions chosen had urban relevance. Data is planned to be released during 2024 jointly with a scientific publication. The municipality is currently using the results as an additional source of information to undergo the urban transformation of this rather small but complex neighborhood \cite{Elisava2022}.

\section{Mental health}

As already mentioned in the introduction, Jusup and coauthors \cite{Jusup2022} define social physics “as a collection of active research topics aiming to resolve societal problems to which scientists with formal training in physics have contributed and continue to contribute substantially”. Societal problems are key drivers in funding schemes from research agencies. Societal problems are intricate, wicked and intrinsically involves uncertainty, complexity and divergence of perspectives \cite{Head2022}. Societal problems require the joint effort of a wide range of actors and stakeholders \cite{Head2022}. 

A recent report \cite{Mazzucato} by the European Research Executive Agency has proposed a mission-oriented research framework where ``bold missions can provide new syntheses that are today impossible and thus will hopefully achieve the breakthroughs that are urgently needed to solve some of the most pressing issues facing our citizens.” The same report states that “citizens can possibly be mobilised to become active participants in missions, for example by cleaning plastics from beaches or by providing real-time monitoring data as enabling technologies develop and become more universally present in society”. Citizen science and social physics may therefore together meaningfully contribute to a societal issue and initiate a research differently. 

We here take one specific societal problem: mental health and the related care provision. The World Health Organisation (WHO) understands mental health as a “state of well-being in which the individual realizes his or her own abilities, can cope with the normal stresses of life, can work productively and fruitfully, and is able to make a contribution to his or her community”. WHO advocates for urgent change in mental health care shifting from a biomedical approach to a recovery model based on principles that include self-determination, resources beyond professional care, and a community approach. WHO also states that ``recovery-oriented care is not about treatment of symptoms but about empowering people to have control of their own lives” and that “we must intensify our collective actions to reform mental health systems towards comprehensive community-based networks of support” \cite{WHO2022}.

Below, we will share two experiences where the participation of people with lived experience in mental health is enhanced and in which the partnership with a civil society organisation becomes totally relevant. These two experiences are summarized in Table \ref{tab:mentalhealth_experiments}. Next subsections will first explain the related digital tools to collect data and the methodologies involved, some of them straightforwardly related to social physics. We will later emphasize those aspects related to the participatory dynamics and the specificities of the research topic.

\begin{table}[t]
\centering
\caption{\label{tab:mentalhealth_experiments}{\bf Citizen science in mental health.} For each research project, we report the non-professional researchers working together with professional scientists, the main research question and its most relevant outcome of the mental health investigations. Columns refer to the context in which the experiments were organised, the digital tool being used to collect behavioural data, the profile of the participants in the digital platforms, the number of participants involved, and the number of decisions collected in each of the experiments. Chatbot number of participants and number of interactions are taken on Dec. 22, 2023 but these numbers are growing everyday.}
\begin{tabular}{lccccc}
\hline \hline
Context & Digital tool & Participant's profile & N. part. & Inter. \\ \hline
Massive event, social & Electronic tablets & Persons with mental & 270 & 3,780  \\
spaces & & health condition, families,\\
& & and social and health-care \\
& & professionals  \\
\multicolumn{5}{l}{Name of the project: Games for Mental Health}\\
\multicolumn{5}{l}{Working group: 20 participants co-designing a behavioural experiment that use social dilemmas.}\\
\multicolumn{5}{l}{Question: Which is the social role of each actor involved in mental health care ecosystem?}\\
\multicolumn{5}{l}{Outcome: Statistical data showing that social workers are the strong ties and that the}\\ 
\multicolumn{5}{l}{participants with lived experience need to play a leading role in the mental health ecosystem}
\\
\\
\hline
Context  & Digital tool & Participant's profile & N. part. & Inter. \\ \hline
Private / personal & Chatbot & Anyone interested  & >766 & >16,711  \\  environments &  & in mental health\\
\multicolumn{5}{l}{Name of the project: CoAct for Mental Health}\\
\multicolumn{5}{l}{Working group: 65 institutional representatives to frame the research, 32 co-researchers involved}\\
\multicolumn{5}{l}{in all research stages, from the project definition to data interpretation and the elaboration}\\
\multicolumn{5}{l}{14 policy recommendations}\\
\multicolumn{5}{l}{Question: What are the key elements of the social support networks in mental health?}\\
\multicolumn{5}{l}{Outcome: Evidence identifying crucial lived experiences within the social support networks and}\\
\multicolumn{5}{l}{the relevance of specific socio-demographic traits}
\\
\\
\hline \hline
\end{tabular}
\end{table}

%Project & Co-researchers & Knowledge Coalition & Citizen Scientists  \\ 
% &  & or Working group  & \\ \hline
%Games  for Mental Health & -- & 20 & 270 \\
%CoAct for Mental Health & 32 & 65 & 704 ?????? \\

\subsection{Digital tools: crowdsourced data collection on mental health and human behaviour}
\label{sect:DMental}

Mental health research reported here looks for behavioural traits organsing a set of experiments to collect data related to these traits. The participants contributes using digital devices through which they can express themselves. Their interaction with digital devices may only take few minutes but it is still possible to get a benefit as a participant. The interactions trigger self-reflection about their own actions during the experiment. Specific personalized reports or updated information about the research progress were also offered. It is however also possible to build mobile-based digital tools that allow for brief but repeated interactions during a longer period of time (several months) as we will also describe below. 

\subsubsection{Social dilemmas in digital platforms}
\label{sec:Public}

\begin{figure}
\centering
\includegraphics[width=1\textwidth]{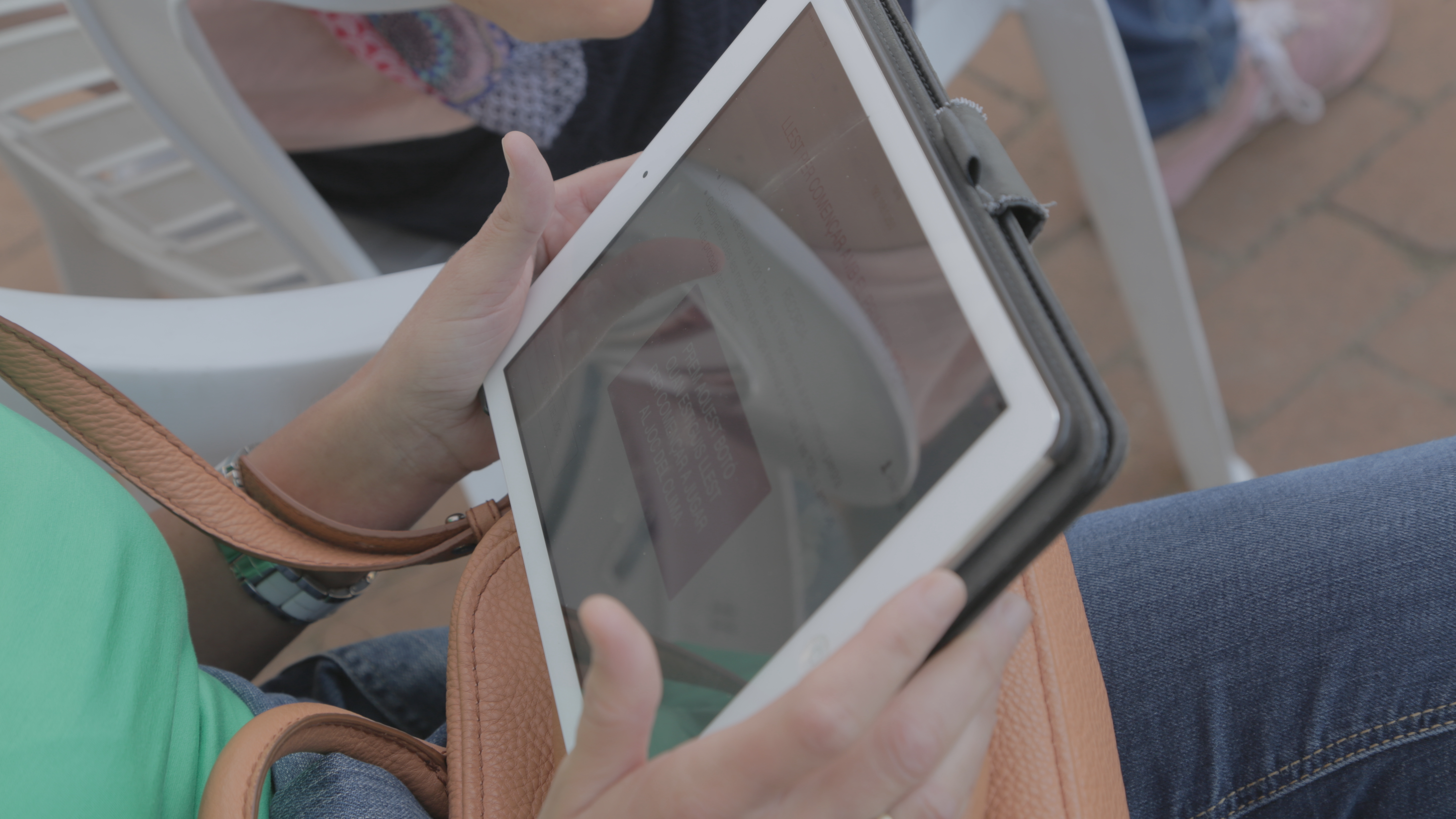}
\caption{{\bf Electronic tablet of Games for Mental Health.} The text of tablet says that not all participants are ready and that the next game is about to start. The Citizen Social Lab digital platform is running through a local server connected to electronic tables with a wifi connection. Participants played in groups of six in an outdoor public space during the 2016 World Mental Health Day related event in Lleida (Spain).}
\label{fig:Games}
\end{figure}

Some of us have recently developed Games for Mental Health (Jocs per la Salut Mental, in Catalan) \cite{Bonhoure2023}. The project (2015-2018) was a joint collaboration with the civil society organisation called Catalonia Mental Health Federation (Federaci\'o Salut Mental Catalunya, in Catalan). The Federation is currently composed of 79 associations of families and users of mental health care services.

As a point of departure, Games for Mental Health took social dilemmas and games which have been extensively explored by the social physics community \cite{Perc2019,Sanchez2018}. The broad aim of the social physics community is to better understand social interactions and human cooperation in a stylized manner within the mental health ecosystem conformed by persons with mental health condition, their families jointly with social and health-care professionals. 

Behavioural experiments have  traditionally been conducted in laboratories generally placed in universities and research centers \cite{Perc2017}. Individuals usually do not know who they are paired with. They take decisions through a computer, in front a screen that provide the instructions to perform the experiment. Also, online experiments has become increasingly popular during last years, particularly through Amazon Mechanical Turk. Online experiments allows to recruit very large samples (thousands of subjects) worldwide and several authors claim that online experiments show similar results to physical laboratory experiments \cite{Perc2017}.

However, neither the physical laboratories nor the online experiments seemed to us the best way to involve the mental health care provision community. We opted to run in-the-field (or in-the-wild) experiments \cite{Perello2012,Sagarra2016}. We relied on our own previous experiences in running several public experiments \cite{Sagarra2016,Vicens2018b} on human behaviour \cite{Gutierrez2014,Gutierrez2016b,Poncela2016,Vicens2018}. This option required to use a flexible digital platform to run the experiment in more natural conditions, where social interactions of the mental health ecosystem are generally taking place as they are the spaces of the mental health care community. Non-permanent and pop-up infrastructure \cite{Sagarra2016} with electronic tablets, a laptop that acted as a local server and a router to set up a local wi-fi network allowed us to run behavioural experiments easily and in a flexible manner. Figure \ref{fig:Games} illustrates these settings with one participant holding an electronic tablet during an experiment in the World Mental Health Day 2016. A specific platform was developed under the name of Citizen Social Lab \cite{Vicens2018b}. The source code developed by one of our collaborators is available at Github \cite{Citizen}. 

\subsubsection{A chatbot to maintain a long-term interaction}

\label{sect:CMental}

\begin{figure}
\centering
\includegraphics[width=1\textwidth]{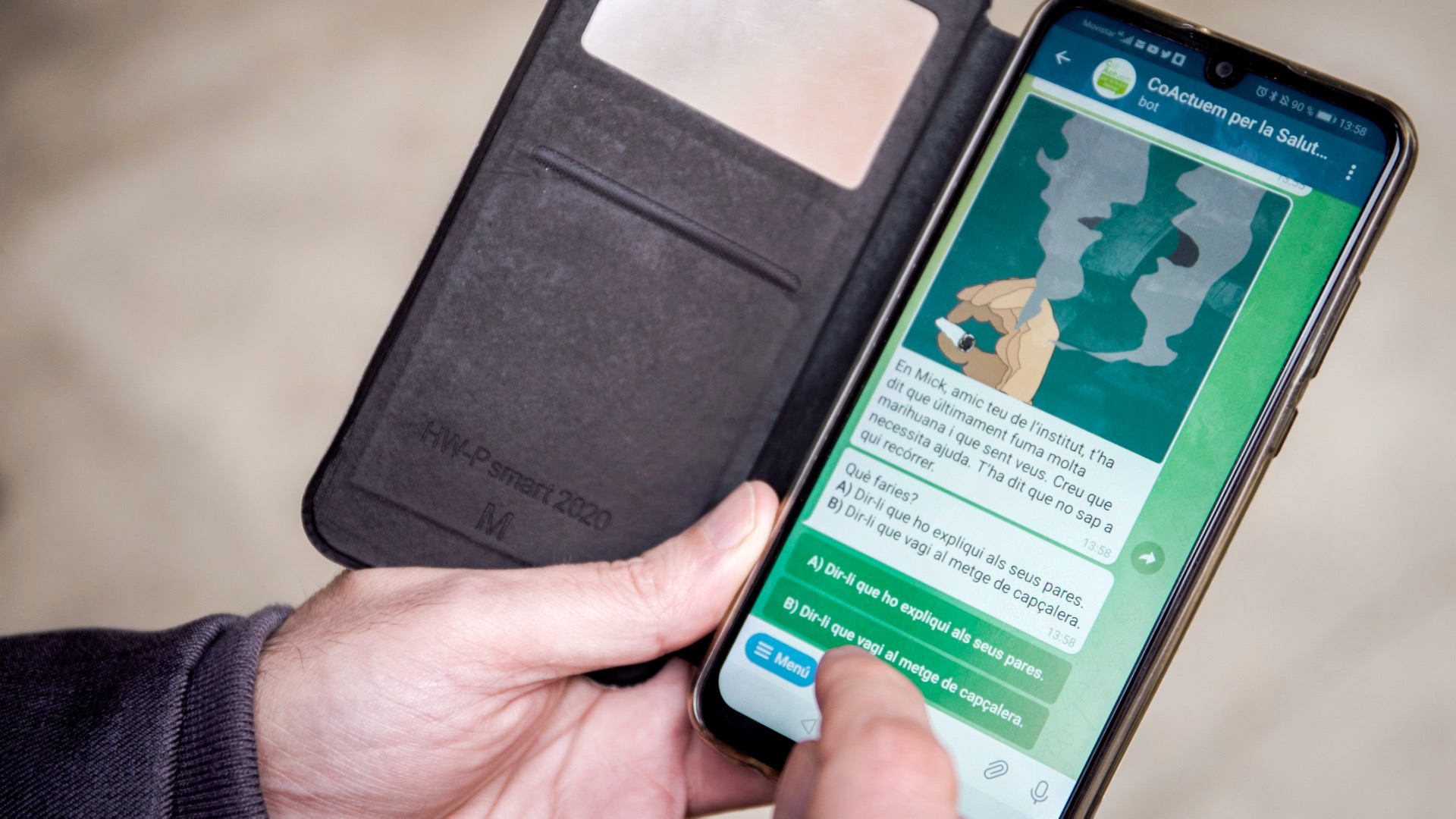}
\caption{{\bf CoActuem per la Salut Mental chatbot.} The mobile of one participant has just received a micro-story from a person with lived experience in mental health. The chatbot is asking a concrete question and the participant needs to select one of the possible answers.}
\label{fig:Chatbot}
\end{figure}

An alternative approach to the event-oriented platforms described in the previous section is to build a chatbot within a popular messaging App. Main motivation is to set up a long-term interaction with the participant. The periodic interactions of the participants from within their different private spaces gives the researchers access to a much more natural context \cite{Bonhoure2023}. We have recently developed the chatbot CoActuem per la Salut Mental \cite{CoActuem} (CoAct for Mental Health, in English) in Telegram. The chatbot is part of the subproject of the Horizon 2020 project titled CoAct \cite{CoAct} (2020-2022). The Federation was also part of the consortia.

The chatbot \cite{Peter2021,CoActgithub} meets high standards of data privacy and is therefore especially apt for exploring any stigmatized social topic, not only mental health. The automated data collection results in well-structured digital data and can be adapted to research questions and scientific methods, especially in the field of social physics, comprising network theory, game theory, epidemics or opinion dynamics \cite{Jusup2022,Barthelemy2019,Castellano2009,Perc2017,Sanchez2018,Wang2016}.

Anyone, with or without a mental health problem, can subscribe to the CoActuem Telegram chatbot. The chatbot automatically sends messages and particularly questions to mutually unconnected participants. The participants receive the contents with a flexible rhythm, e.g. daily, and can decide on the timing of their answering themselves. Figure \ref{fig:Chatbot} show the chatbot and the messages that a participant receives in its own mobile device.

First dialogue that the participant subscribed to the chatbot receives welcomes them and ask for providing informed consent. The second dialogue serves as capacitation building and gives examples of what later will be the two types of dialogues the participants are asked to answer. The participants is also asked to complete an extensive socio-demographic survey, comprising 32 questions.

After these interactions, the participant periodically receives a dialogue which includes a micro-story experienced and put into words by persons with lived experience in mental health (see co-creative sessions below reported in Section \ref{Sect:SocialSupport}). One type of micro-story, “share experiences”, asks the participants whether they lived the shared experience, too, and whether someone in their surrounding lived it. The other type, “find solutions”, asks the participants to decide between two different reactions to the presented micro-story, also written down by persons with lived experience in mental health. All in all, participants can react to up to 222 microstories and receive a total of more than 300 dialogues. The source code is available at Github \cite{CoActgithub} and is also documented in Zenodo \cite{Peter2021}. 

\begin{figure}
\centering
\includegraphics[width=1\textwidth]{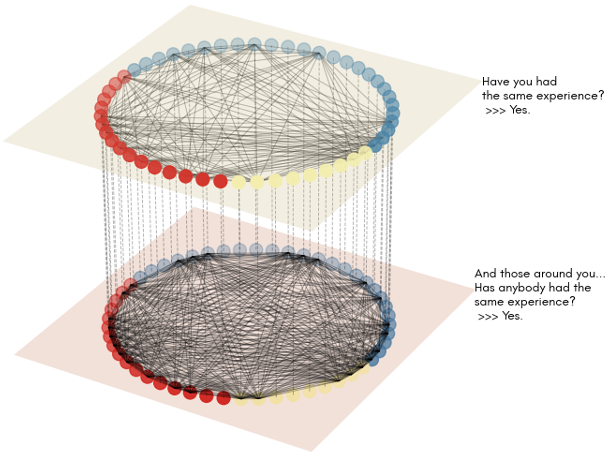}
\caption{{\bf Network construction of a sample of participants in the chatbot.} Each node represents a participant (women in blue, men in red, other answers in yellow). In each layer, links can be built if two participants have provided the same answer to the same micro-story. Top layer corresponds to the question whether they have had the same experience and most crowded layer correspond to the question if anybody around them have had the same experience.}
\label{fig:coact_network}
\end{figure}

The interaction via buttons facilitates the later analysis considerably. All sent contents and data collected from the participants are stored in a database in a way that is easily accessible during its runtime. After having collected the participants’ answers and for scientific research, the unstructured data is converted into several tables that grant quick access to the data analysts. One possible analysis is to build a network of micro-stories of type “share experiences”. A link can be built based on whether a pair of participants answers in the same manner different micro-stories. It is then possible to build a network. The nodes can be taking information from the socio-demographic survey (e.g. gender self-reported identity). We exemplify these possibilities in Figure \ref{fig:coact_network} where we select a group of participants and build a multiplex network with the two answers from each participant to a set of micro-stories of type “share experiences”.

\subsection{Participation: lived experiences in mental health}
\label{sect:PMental}

WHO is emphasizing that to transform mental health care provision it is necessary to enhance the participation of people with lived experience in mental health \cite{WHO2022}. There are also other several official recommendations to ensure participation of persons with mental health problems and their families, at all levels, including research, design, and implementation of services and programmes \cite{EC2016,UN2006}. 

A citizen science project in this context can therefore provide new tools and strategies to valorize inside a scientific research the knowledge from the people with lived experience. This expertise is mostly held at an individual level and it is barely socialized due to social stigma or the limited number of spaces available to put knowledge in common. The scientific research in this community-based frame is still limited and there is a lack of data to provide evidence on the benefits and challenges behind the community-based mental care provision \cite{Bonhoure2023}.

\subsubsection{Behavioural traits within the mental health care community}

With the support of the Federation, a working group was created to co-design the experiments. The working group was formed by a range of 20 people with diverse experiences and expertise: people with mental health conditions, non-professional caregivers, relatives, social workers, mental health nurses, psychologists, and psychiatrists, as well as experts and board members of the Federation. During a set of workshops, we together decided which were the most relevant behavioural traits (cooperation, trust, reciprocity and sense of collectiveness) and the most interesting games (Prisoner's Dilemma, Trust Game and Collective Risk Dilemma). The working group also framed the experimental settings and the methods and protocols related to see the experiments also as an experiential activity that enables participants to self-reflect about their position within the mental health care community and in the mental health ecosystem.

As carefully described in Ref. \cite{Bonhoure2023}, we first ran the Games for Mental Health (120 participants) as one of the activities in a mass event at a regional level during the week of the World Mental Health Day. Participants belonged to the mental health associative movement, including people with mental health conditions, their families, and social and health-care professionals from the sector. We also embedded the experiment in three social spaces. In these settings, 150 more people played the games, thus reaching out a total number of participants of 270 individuals. This number of participants is below the number of experimental subjects gathered in Mechanical Turk behavioural experiments but the numbers result to be comparable to those gathered in physical laboratories that have studied mental conditions in clinical settings \cite{Bonhoure2023}.

Results obtained from the public experiments with the active and conscious participation people with lived experience in mental health, non-professional caregivers (e.g., family members) and professional caregivers (e.g., social workers) was published in a scientific journal \cite{Cigarini2018} and the experimental data was also released \cite{Cigarini2018c}. Results showed that caregivers are the strong ties in the community as they showed high levels of cooperation and optimism. Also, participants with lived experience also showed to play a leading role in the ecosystem because they were putting larger efforts towards reaching a collective goal in an specific game. The study also emphasized the need to take complex systems science perspective when looking at the community involved in mental health care provision. A press conference was also jointly organised by the Universitat de Barcelona and the Federation. An easy-to-read document with policy recommendations was also released jointly with the press release of the scientific paper \cite{Cigarini2018b}.

\subsubsection{The social support networks in mental health}
\label{Sect:SocialSupport}

\begin{figure}
\centering
\includegraphics[width=1\textwidth]{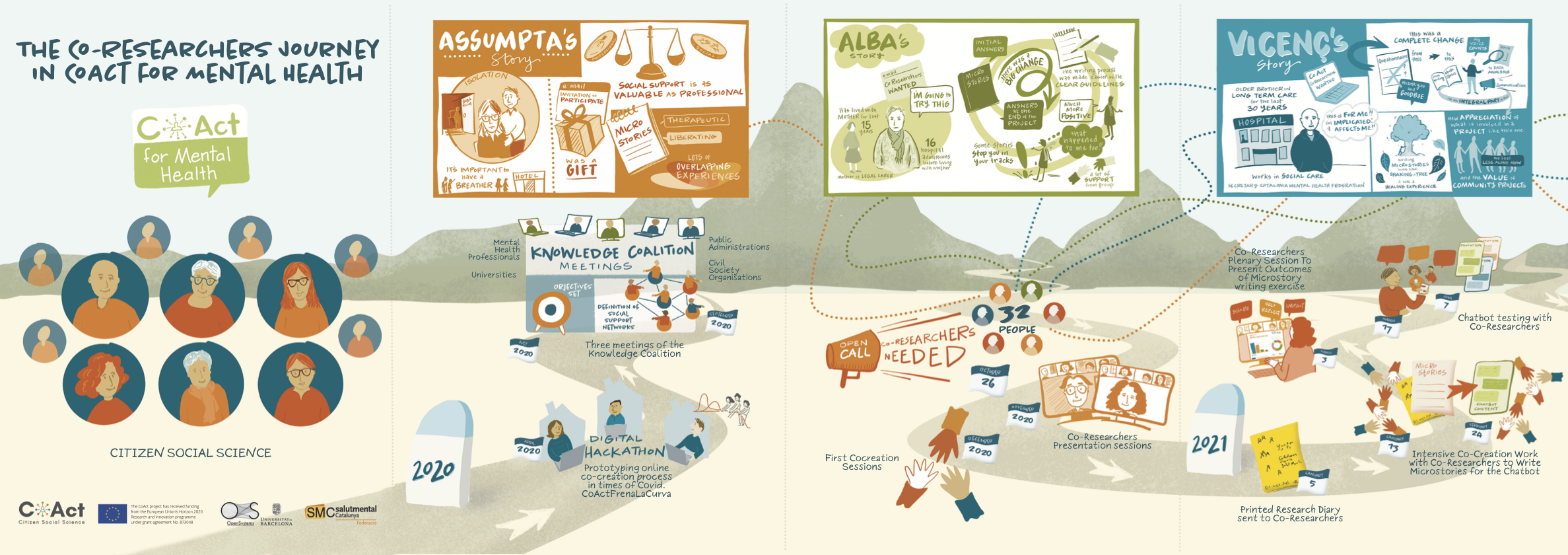}
\includegraphics[width=1\textwidth]{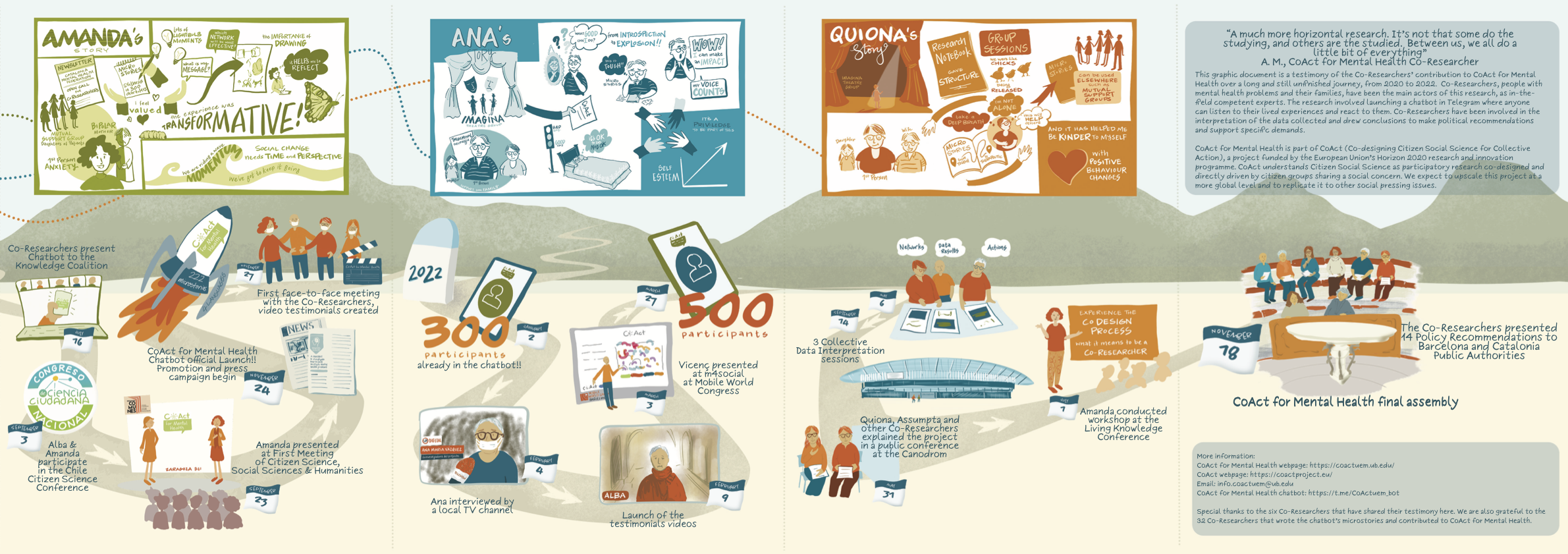}
\caption{{\bf The co-researchers journey in CoAct for Mental Health.} We collect the direct experience of 6 co-researchers that participated in all phases of the research on social support networks in mental health. We here cut in two slice the long journey made between 2020--2022. The pdf file is available in Zenodo \cite{Harrison2022}. Graphic document made by Verity Harrison.}
\label{fig:journey}
\end{figure}

Social physics and complex systems science can also find of interest the so-called social support networks in mental health. These networks are a broad construct of social resources that an individual perceives \cite{Turner2009}. They also refer to mutual assistance, guidance, and validation about life experiences and decisions \cite{Zhou2014}. However, there is no much data to further understand how social support networks work in the context of mental health. This lack of empirical data is hindering the transforming mental health care provision towards an enhancement community-based health services. 

With the high ambition of maximizing the collaboration with citizens, a group of 32 co-researchers (either persons with mental health issues or their family members) were gathered in the earliest stage of the project. They were part of the research team as they were considered competent experts in-the-field, based on their daily lived experiences. This hand-in-hand contribution was parallel to another effort in creating a community that we gave the name of knowledge coalition. It was formed by 65 institutional representatives from public administrations, civil society organisations, educational organisations, and academia working on mental health. 

The first step made was to together frame the research. Collaborative documents and definitions were also prepared, to establish a common definition of social support network and a shared basis on the research approach being taken. Then, co-researchers were invited to write personal stories about their lived experiences of social support, accompanied by a professional writer and a professional graphic artist who illustrated the stories. The process of writing microstories was not a straightforward task and a Research Diary was produced to give support to co-researchers. The Research Diary \cite{Peter2021} was meant to be used as physical supporting material after, during and between the sessions to favour individual and self-reflective processes. The resulting micro-stories were then shared to anyone subscribed in the CoActuem per la Salut Mental chatbot as described in Section \ref{sect:CMental}. In this way, we wanted to explore whether participants in the chatbot and people around them have lived similar experiences. 

We together hypothesized that this serve us as a first proxy to explore the roles of everyone and the experiences that can help to better identify key aspects of social support networks (see also Figure \ref{fig:coact_network}). This is still an ongoing work with the co-researchers and the knowledge coalition. However, it has already served to support with evidences a set of 14 policy recommendations delivered to local public authorities (Barcelona City Council and commissioner for the National Mental Health Plan at a regional level). The document under the form of a policy brief was discussed during an assembly (November 2023) with knowledge coalition members and co-researchers \cite{Mitats2022,Mitats2023}. The event was also open to anyone who has participated in the chatbot or that was simply interested in mental health.

Both the co-researchers and the knowledge coalition has been involved for a quite long term period (2-3 years) and for a wide variety tasks within a research which is indeed not fully ended. Figure \ref{fig:journey} describes the co-researchers’ contributions so far under the form of a journey. The same dynamics can be used to any other citizen science project facing a societal problem where a group of co-researchers can contribute with their lived experiences. We are still under the process of writing several related scientific publications which some of them will be co-authored by some co-researchers.
 
\section{Discussion and Conclusion}\label{sec13}

We have aimed to call attention to some citizen science practices and encourage social physics community to adopt them and expand them. We believe that citizen science offers ways to circumvent current obstacles to the further advancement of social physics and by extension the broader field of computational social science \cite{Verhulst2023,Leslie2023}. Experts have observed inadequate data sharing paradigms to consolidate scientific outcomes and the absence of clear mechanisms to perform ethical research \cite{Lazer2020,Nature2021,Verhulst2023,Leslie2023}. Running an ethical research is not only related to privacy issues and regulation compliance. It also refers to a further consideration socio-economic inequalities and vulnerabilities as part of the research agenda, to the inclusion of concerned citizens collectives and eventually those in a vulnerable position in research activities,  or to an increment of the citizens’ sense of ownership on their own digital data \cite{Lazer2020,Sadowski2021}.

Along these lines, citizen science practices can become a very relevant approach. In citizen science, participants takes an active role in the research definition, the research question, the results interpretation or the transformation of the scientific results into valuable knowledge \cite{CoAct}. If we talk about social phenomena, both the motivation and the value of the co-produced knowledge are deeply involved in participants’ everyday life and the results can in turn influence policies and actions to promote social change \cite{Albert2021,Bonhoure2023,Schade2021}. In a broader level, benefits also cover a wide list of aspects that surpass scientific interest and include innovative STEM or STEAM education \cite{Senabre2018,Roche2020}, democratic values enhancement \cite{Pearse2020}, social inclusion promotion \cite{Paleco2021,Senabre2021}, or evidence based policy making \cite{Bonhoure2023,Schade2021,Criscuolo2022,Sauermann2020}. Furthermore, citizen science practices can be key in the problem identification and agenda setting in scientific research related to societal problems \cite{Jusup2022} or become a key element in sustainability transitions \cite{Criscuolo2022,Sauermann2020}. And, both globally and locally, citizen science projects can bring to the table non-traditional data sources to contribute to the United Nations Sustainable Development Goals \cite{Fritz2021}.

We have here focussed in our own experience in two particular topics which can be linked to the approaches and methods that social physics typically use \cite{Jusup2022}. Learnings from the shared experiences being presented in this paper are many. We would like to stress the importance of building new digital tools which need to be versatile and adaptive to participants' needs, concerns and constraints. This generally leads to the use of electronic tablets and mobile devices which are more accessible and closer to daily-life activities of the participants. Sometimes, it is however more effective to customize existing digital tools rather than building new ones. Customization may involve new features and uses but most importantly it allows to run in-the-field human behaviour experiments in more natural conditions. These efforts allows to collect data avoiding the intermediation of tech companies or to get data much more aligned to concerned groups of citizens. In both cases, these data can be complementary to existing data. The data can allow to narrow down the research question and more effectively address very specific issues. 

Also, participatory dynamics needs to be carefully organised. One can take windows of opportunity in massive events such as a science festival or the public events during the World Mental Health Day. It is however also important to approach to several contexts much closer to daily social interactions, even if it is for few minutes or for a much longer period of time. In any case, it is particularly relevant to stablish a long-lasting relationship with specific groups that have a clear motivation (young students or a group of old women). Building a participatory research with a civil society organisation can also make the whole effort much more robust as research can be better framed. Knowledge being co-produced have then more chances to become actionable in terms of policies and collective actions. These collaborations are valuable to construct communities around the research (such as the knowledge coalition described in CoAct for Mental Health) but it is even more important to incorporate as co-researchers persons with lived experience as they already have invaluable knowledge in their hands. Further reflections of a citizen science for social physics can be highly benefit from ongoing discussion around the so-called citizen social science or social citizen science \cite{Albert2021,Bonhoure2023}. Citizen social science enhances the social dimension in citizen science and can be understood as participatory research co-designed and directly driven by citizen groups sharing a social concern \cite{Bonhoure2023}. 

Several other topics that social physics investigates can also be part of these ongoing efforts. Many techniques from social physics can be further developed or even reformulated when considering citizen science data. For instance, GPS collected data can help to build completely new stochastic models with special interest in human  micro-mobility \cite{Larroya2023}. Social interactions can also be characterized with the definition of specific traits that quantifies ties among individuals within the mental health care ecosystem \cite{Cigarini2018}. The sample of participants in citizen science projects might be more limited (hundreds of participants) compared to other efforts coming from social physics, complex systems and computational social science. However, the smaller size of the datasets is compensated with the fact that citizen science can gather richer data or at least more meaningful data. The collected data can be more oriented to specific research questions. In the case of pedestrian mobility, the location data is from specific socio-demographic profiles and belongs to already identified purposed-based mobility or narrows down the analysis to a concrete neighbourhood. In the cases related to mental health care, it is possible to run behavioural experiments to characterize the social interactions in terms of traits such as cooperation or trust and reinterpret these traits to better understand the mental health ecosystem. Or alternatively, it is possible to map out and better understand key elements in social support networks based on lived experiences narrated under the form of micro-stories with complex networks, clustering analysis, and machine learning algorithms.

Overall, the adoption of a citizen science for social physics increases research legitimacy and the potential to transform scientific knowledge into specific actions and policies. Crowd-sourced citizen science practices around complex social phenomena extend the research effort to the general public. A deep involvement of co-researchers and knowledge coalition members in scientific research can promote an evidence-based culture beyond the academic context and can open new deliberative spaces to a wide variety of groups and organisations in our societies. This joint effort has the potential to nurture richer public debates around many crucial societal challenges.

\backmatter

\bmhead{Acknowledgments}

We acknowledge the participation of all volunteers involved and all collaborators and former members of the group OpenSystems. We particularly thank the Consorci d’Educació de Barcelona and the Barcelona City Council through its Citizen Science Office and Salut Mental Catalunya for their commitment to the projects reported and their support to citizen science practices. This work was partially supported by Ministerio de Ciencia e Innovación (MCIN, Spain), Agencia Estatal de Investigación (AEI) AEI/10.13039/501100011033 and Fondo Europeo de Desarrollo Regional (FEDER) [grant number PID2019-106811GB-C33, JP, FL, IB and FP]; by the ERA-Net Urban Transformation Capacities (ENUTC) program [OPUSH, contract number 101003758, JP, FL, IB and FP] and by MCIN/AEI/10.13039/501100011033 and European Union NextGenerationEU/PRTR [grant number PCI2022-132996, JP, FL, IB and FP]; by Horizon 2020 program [COACT, contract number 873048, JP, FL and IB]; by Horizon Europe WIDERA program [SENSE., contract number 101058507, JP]; by Generalitat de Catalunya (Spain) through Complexity Lab Barcelona [grant numbers 2017 SGR 608 and 2021 SGR 00856; JP, FL, IB and FP].

\section*{Statements and Declarations}
\bmhead{Competing Interests} The authors declare no competing interests.

%\section*{References}

\end{document}